\documentclass[pra,twocolumn,showpacs,preprintnumbers,amsmath,amssymb]{revtex4-1}

\usepackage{graphicx}
\usepackage{dcolumn}
\usepackage{bm}
\usepackage{psfrag}
\usepackage{epsfig}
\usepackage{amsmath}
\usepackage{amssymb}
\usepackage{color}
\usepackage[FIGTOPCAP,raggedright,nooneline]{subfigure}

\newcommand{\bra}[1]{\langle #1 |}
\newcommand{\ket}[1]{| #1 \rangle}

\newcommand{\nn}{\nonumber}
\newcommand{\Bra}{\langle}
\newcommand{\Ket}{\rangle}

\newcommand{\ignore}[1]{}

\begin{document}

\title{Signatures of single site addressability in resonance fluorescence spectra}

\author{Peter Degenfeld-Schonburg}
\email{peter.degenfeld-schonburg@ph.tum.de}
\author{Elena del Valle}
\author{Michael J. Hartmann}
\affiliation{Technische Universit{\"a}t M{\"u}nchen, Physik Department I,
James Franck Str., 85748 Garching, Germany}

\date{\today}

\begin{abstract}
Pioneering methods in recent optical lattice experiments allow to focus laser beams down to a spot size that is 
comparable to the lattice constant. Inspired by this achievement, we examine the resonance fluorescence spectra of 
two-level atoms positioned in adjacent lattice sites and compare the case where the laser hits only one atom 
(single site addressing) with cases where several atoms are illuminated. In contrast to the
case where the laser hits several atoms, the spectrum for single site addressing is no longer symmetric 
around the laser frequency. The shape of the spectrum of fluorescent light can therefore serve as a test for 
single site addressing. The effects we find can be attributed to a dipole-dipole
interaction between the atoms due to mutual exchange of photons.
\end{abstract}

\pacs{42.50.Nn,37.10.Jk,42.25.Fx}
\maketitle

%
%
\section{Introduction}
Over the course of the last decades, the exploration of the radiative properties of
laser driven atomic systems advanced at a stunning pace in the field of quantum
optics. The fluorescence light of a coherently driven two-level atom is a common example
in most textbooks \cite{WMbook,Scullybook}. For a coherent laser drive, the predicted Mollow spectrum is a symmetrical 
three peak spectrum with center-side band separation given by the Rabi frequency and the detuning of the driving laser 
field \cite{Mollow, Elena1, Elena2}. Taking more than one atom into account, other interesting
features of the spectrum arise due to effects of coherent and incoherent inter-atomic interactions \cite{Ficek}. 
The question to what extend the fluorescence spectrum is altered in the presence of
collective effects is thus of great interest since it contains informations about
the physical setup of the atomic system \cite{Scully}.

In this work, we clarify how the shape of the spectrum alters when a single atom within the atomic ensemble
is addressed by an external driving field. Therefore we compare the spectra of the situation where a laser 
illuminates all atoms with the situation where the laser illuminates only one atom. The usual symmetry of the 
spectrum breaks down if the distance between neighboring atoms is such that their dipole-dipole interaction
via mutual exchange of photons is comparable to the magnitude 
of the driving strength.

Our investigation of the resonance fluorescence spectrum under the assumption that single atoms in an atomic 
ensemble can be addressed by a laser beam, is motivated by recent experiments with optical lattices 
\cite{greiner, bloch} in a Hubbard-regime that
trap neutral atoms in the lattice sites. High resolution imaging systems with an optical resolution of about 
600-700 nm allow to resolve the fluorescent light emitted from the atoms in individual lattice sites of a two 
dimensional lattice. Hence, a laser beam traveling on the same path just like the fluorescence light going through
 the imaging system, but in reverse direction, can be focused onto a single site or single atom with a full-width 
at half-maximum (FWHM) of again 600-700 nm~\cite{bloch2}. Such single site addressing allows to investigate local
properties of quantum many-particle systems~\cite{MH1,MH2}. 

The origin of the effects we find for the fluorescence spectra of neutral atoms in optical lattices lies in the
typical inter-atomic separation. On the one hand, the atoms are far enough from each other to open up the possibility 
of single atom addressing by focused laser beams. On the other hand, the separation
is small enough so that an inter-atomic coupling mediated by mutual exchange of scattered photons still influences 
the collective behavior. Otherwise the fluorescence spectrum would not differ from the well known Mollow spectrum \cite{Mollow}.

The remainder of the paper is organized as follows.
In section \ref{sec_master_ham}, we set up a master equation that describes the coupling of $N$ two-level atoms to the quantized 
electromagnetic field. Due to small inter-atomic separations we have to account for an effective atom-atom coupling
mediated by the quantized field. 
The effective atom-atom coupling, namely the dipole-dipole interaction and the collective
damping rate, will be discussed in detail. We also define our notion of single site addressing 
motivated by the experiment~\cite{bloch2}.
As our investigations focus on the fluorescence spectrum of the atomic ensemble, the power spectrum will be introduced in
section \ref{power_spectrum}. We then present our numerical results in section \ref{results} for a number of up to five 
two-level atoms representing single site addressability in a 1D or 2D optical lattice. 
In all cases we find a broken symmetry in the fluorescence spectra if only one atom is addressed
by the laser field. We demonstrate that this effect only occurs if the dipole-dipole interaction is finite.
In section \ref{symmetry chapter} we make some general remarks and provide conclusions about symmetric power spectra.
In section \ref{exp} we state possible experimental applications to test single site addressability with 
resonance fluorescence measurements. Finally we give our conclusions in section \ref{conclusionsattheend}.

\section{Hamiltonian and master equation} \label{sec_master_ham}

In our model we consider $N$ identical two-level atoms at fixed positions $\textbf{r}_\mu$ and define the 
distances $\textbf{R}_{\mu\nu}=\textbf{r}_\mu-\textbf{r}_\nu$. We demand, however, that all atoms lie inside 
a two-dimensional plane to which we will refer to as the atomic plane. The ground state of the atom $\mu$ is denoted 
by $\ket{g_\mu}$ and the excited state by $\ket{e_\mu}$, where $\mu=1,2,...,N$ labels the atoms. 
Apart from the laser that illuminates them, the atoms couple to all modes of the quantized electromagnetic 
vacuum. The time evolution of the atomic system can therefore be described by the master equation~\cite{kiffner},
\begin{align} \label{eq:master}
& \partial_{t}\rho(t)=\mathbf{L}(\rho(t)) =-\frac{i}{\hbar}\left[H,\rho(t)\right] \nn \\
& + \frac{\gamma}{2}\sum_{\mu=1}^N\left(2\sigma_{\mu}^{-}\rho(t)\sigma_{\mu}^{+}-\sigma_{\mu}^{+}
\sigma_{\mu}^{-}\rho(t)-\rho(t)\sigma_{\mu}^{+}\sigma_{\mu}^{-}\right) \nn \\
& +\sum\limits_{\genfrac{}{}{0pt}{2}{\mu,\nu=1}{\mu\not=\nu}}^N\Gamma^{\mu\nu}\left(2\sigma_{\nu}^{-}
\rho(t)\sigma_{\mu}^{+}-\sigma_{\mu}^{+}\sigma_{\nu}^{-}\rho(t)-\rho(t)\sigma_{\mu}^{+}
\sigma_{\nu}^{-}\right),
\end{align}
with Liouvillian $\mathbf{L}(\cdot)$, where the operators $\sigma_\mu^+=\ket{e_\mu}\bra{g_\mu}$ $\left(\sigma_\mu^-=\ket{g_\mu}\bra{e_\mu}\right)$ create 
(destroy) an excitation in the $\mu$-th atom. The decay rate $\gamma$ accounts for the spontaneous emission 
from a single atom, while the collective damping rates $\Gamma^{\mu\nu}=\Gamma^{\nu\mu}$ account for the decay of 
the collective atomic states. Since the spatial separation of the atoms in our setup is about 1/2 of the wavelength 
corresponding to the atomic transition, the often used assumption of independent quantum environments~\cite{MH3,MH4} for the 
individual atoms is in our study not justified. Consequently the collective decay processes associated to 
$\Gamma^{\mu\nu}$ are indeed important~\cite{Ritsch}. This is also confirmed by our results. 

\begin{figure}[tbp]
\includegraphics[width=8cm]{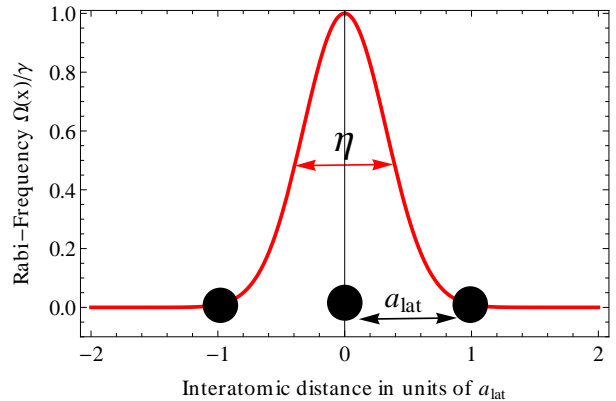}
\caption{\label{laser_focus} Notion of single site addressing in an optical lattice. The atoms are fixed in 
the lattice sites while a laser beam traveling perpendicular to the inter-atomic axis is focused onto one site. 
The beam profile is given by a Gaussian envelope with a FWHM $\eta$ as given in Eq.~(\ref{eq:gaussian_beam}). 
The parameters $a_l$ and $\eta$ match the experimental parameters of Ref. \cite{greiner}.}
\end{figure}

The unitary part of the time evolution of the reduced density matrix $\rho(t)$ is governed by the Hamiltonian
\begin{equation} \label{eq:ham}
H=H_0+H_L+H_{dd},
\end{equation}
consisting of three parts; the bare atomic part $H_0$, the laser-atom coupling $H_L$ and the dipole-dipole 
interaction potential $H_{dd}$. In a frame that rotates with the frequency $\omega_L$ of the driving 
laser the bare Hamiltonian
\begin{equation} \label{eq:barehamilton}
H_0=\frac{\hbar\Delta}{2}\sum_{\mu=1}^N\sigma_{\mu}^z=\frac{\hbar\Delta}{2}\sum_{\mu=1}^N
\left(\ket{e_\mu}\bra{e_\mu}-\ket{g_\mu}\bra{g_\mu}\right)
\end{equation}
is determined by the detuning $\Delta=\omega_0-\omega_L$, where $\omega_0$ is the frequency of the atomic 
transition. The laser-atom coupling in turn depends on the Rabi frequency
\begin{equation}
H_L=\hbar\sum_{\mu=1}^N\Omega_\mu(\sigma_\mu^++\sigma_\mu^-).
\end{equation}
The atoms are driven by a laser traveling perpendicular to the atomic plane. Hence, the laser field at each 
atom has the same phase. Without loss of generality we thus choose $\Omega_{\mu}$ to be real and positive 
for all atoms. We consider a Gaussian beam profile focused onto the atomic plane and model the field amplitude 
at the position of the $\mu$-th atom by the Rabi frequency
\begin{equation}\label{eq:gaussian_beam}
\Omega_\mu\equiv\Omega(\textbf{r}_\mu)=\Omega_0 e^{-4\ln2\cdotp \frac{\textbf{r}_\mu^2}{\eta^2}},
\end{equation}
where $\Omega_0\in \mathbb{R}$ represents the intensity at the center of the beam, which defines the point 
of origin, with a FWHM of $\eta$, see Fig.~\ref{laser_focus}.
This definition is consistent with our notion of single site addressing, pictured in Fig.~\ref{laser_focus},
which is motivated by the optical lattice 
experiments \cite{bloch,greiner,bloch2} with single atom occupation on one lattice site and where the 
central atom is addressed by a focused Gaussian laser beam. In our model 
we assume the atoms to be stationary during the illumination process, i.e. we consider a lattice in 
the Mott Insulator regime.

\begin{figure}[tbp]
\includegraphics[width=8cm]{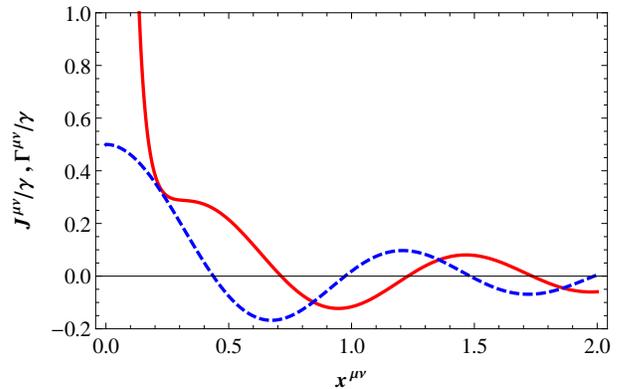}
\caption{\label{collectiv_parameters} (Color online) The dipole-dipole interaction $J^{\mu\nu}$ (solid red line) and the collective 
damping rate $\Gamma^{\mu\nu}$ (blue dashed line) in units of $\gamma$ as a function of the 
inter-atomic separation $x^{\mu\nu}$ with $\alpha^{\mu\nu}=\frac{\pi}{2}$.}
\end{figure}

The coupling to the electromagnetic vacuum induces a dipole-dipole interaction between the atoms due to mutual
exchange of photons.
This dipole-dipole interaction arises in the derivation process of the master equation. 
Just as the collective damping rate it is a consequence of the fact that all atoms couple to 
the same quantum environment.
The dipole-dipole interaction $J^{\mu\nu}$, which should not be confused with the interaction between 
two static dipoles, is the coherent counterpart to the incoherent collective damping rate $\Gamma^{\mu\nu}$.
The operator form of the dipole-dipole Hamiltonian can be taken to read
\begin{equation} \label{eq:ddhamilton}
H_{dd}=\hbar \sum\limits_{\genfrac{}{}{0pt}{2}{\mu,\nu=1}{\mu\not=\nu}}^N\frac{J^{\mu\nu}}{2}
\left(\sigma_{\mu}^{+}\sigma_{\nu}^{-}+\sigma_{\nu}^{+}\sigma_{\mu}^{-}\right),
\end{equation}
with interaction strength $J^{\mu\nu}=J^{\nu\mu}$,
which depends on only two parameters. Namely, the distance $\rvert\textbf{R}_{\mu\nu}\rvert$ 
between the atoms and the 
angle $\alpha^{\mu\nu}=\measuredangle (\textbf{R}_{\mu\nu},\vec{d}_0)$ between the inter-atomic separation 
vector and the dipole moment $\vec{d}_0$ of the atoms. By introducing the dimensionless 
quantity $x^{\mu\nu}:=\frac{\rvert\textbf{R}_{\mu\nu}\rvert}{\lambda_0}$, where $\lambda_0$ is 
the wavelength corresponding to the energy splitting $\omega_0$, we find the following form for the 
dipole-dipole interaction
\begin{equation}\label{eq:DDint}
\begin{split}
&J^{\mu\nu}(\alpha^{\mu\nu},x^{\mu\nu})=\frac{3}{4}\gamma\left\{[\cos^2(\alpha^{\mu\nu})-1]\frac{\cos(2\pi x^{\mu\nu})}{2\pi x^{\mu\nu}}\right.\\
& \left. +\left[1-3\cos^2(\alpha^{\mu\nu})\right]\left[\frac{\sin(2\pi x^{\mu\nu})}{(2\pi x^{\mu\nu})^2}+\frac{\cos(2\pi x^{\mu\nu})}{(2\pi x^{\mu\nu})^3}\right]\right\},
\end{split}
\end{equation}
and for the collective damping rate
\begin{equation}
\begin{split}
&\Gamma^{\mu\nu}(\alpha^{\mu\nu},x^{\mu\nu})=\frac{3}{4}\gamma\left\{\left[1-\cos^2(\alpha^{\mu\nu})\right]\frac{\sin(2\pi x^{\mu\nu})}{2\pi x^{\mu\nu}}\right.\\
& \left. +\left[1-3\cos^2(\alpha^{\mu\nu})\right]\left[\frac{\cos(2\pi x^{\mu\nu})}{(2\pi x^{\mu\nu})^2}-\frac{\sin(2\pi x^{\mu\nu})}{(2\pi x^{\mu\nu})^3}\right]\right\},
\end{split}
\end{equation}
given in units of the spontaneous 
emission rate $\gamma =\frac{\omega_{0}^3|\vec{d}_0|^2}{3\pi\varepsilon_{0}\hbar c^3}$.
Both quantities decay asymptotically as $\sim\frac{1}{x^{\mu\nu}}$, see Fig.~\ref{collectiv_parameters}, 
which indicates that the spectral properties of many atoms separated by a large distance do not differ from the 
spectral properties of a single atom. In order to find 
collective dynamics the inter-atomic distances should be at least of the order of the atomic wavelength, 
i.e. $x^{\mu\nu}\lesssim 1$. On these small length scales addressing of a single atom becomes challenging. 
We show that the collective quantities will affect the spectral response of the atoms and reveal whether 
single atom addressing takes place or not.

\section{Spectrum of Resonance Fluorescence} \label{power_spectrum}

In the following we focus on the spectral distribution of the fluorescence light emitted 
by the atoms in the steady state limit. First we concentrate on the total steady state intensity which is given 
by the normally ordered one-time correlation function of the emitted electric field \cite{Glauber},
\begin{equation}\label{eq:IntSS}
I_{ss}(\textbf{r})=\lim_{t\rightarrow\infty}
\langle\hat{\textbf{E}}^{(-)}(\textbf{r},t)\cdot\hat{\textbf{E}}^{(+)}(\textbf{r},t)\rangle.
\end{equation}
Here, $\hat{\textbf{E}}^{(+)}$ $(\hat{\textbf{E}}^{(-)})$ denotes the positive (negative) frequency 
part of the field operator, which is related to the atomic transition operator $\sigma_\mu^-$ by \cite{Lehmberg},
\begin{equation}\label{sourceterm}
\hat{\textbf{E}}^{(+)}(\textbf{r},\hat{t})=
-\frac{\omega_0^2 \hat{\textbf{r}}\times(\hat{\textbf{r}}\times\vec{d}_0)}{4\pi\varepsilon_{0}c^2r}
\sum_{\mu=1}^N\sigma_\mu^-(\hat t)e^{-ik_0\hat{\textbf{r}}\cdotp \textbf{r}_\mu},
\end{equation}
with the retarded time $\hat{t}=t-\frac{r}{c}$ at a point $\textbf{r} = r\hat{\textbf{r}}$ in the far-field zone. 
The physical picture here is that every photon which is 
annihilated in the detection process had to be emitted by an atom at an earlier time $\hat{t}$. 
Turning back to the spectral properties of the atoms we introduce the so called power spectrum. 
It displays the emitted fluorescence intensity per energy interval
and is given by the Fourier transform of the two-time correlation function of the electric field,
\begin{equation}
S_{ss}(\textbf{r},\omega) =\lim_{t\rightarrow\infty}\int_{-\infty}^{\infty} 
\frac{d\tau}{2\pi}e^{-i\omega \tau}
\langle\hat{\textbf{E}}^{(-)}(\textbf{r},t+\tau)\cdot\hat{\textbf{E}}^{(+)}(\textbf{r},t)\rangle.
\end{equation}
\begin{figure*}[htbp]
\subfigure[{}]{\label{twoatomspeca}\includegraphics[width=8cm]{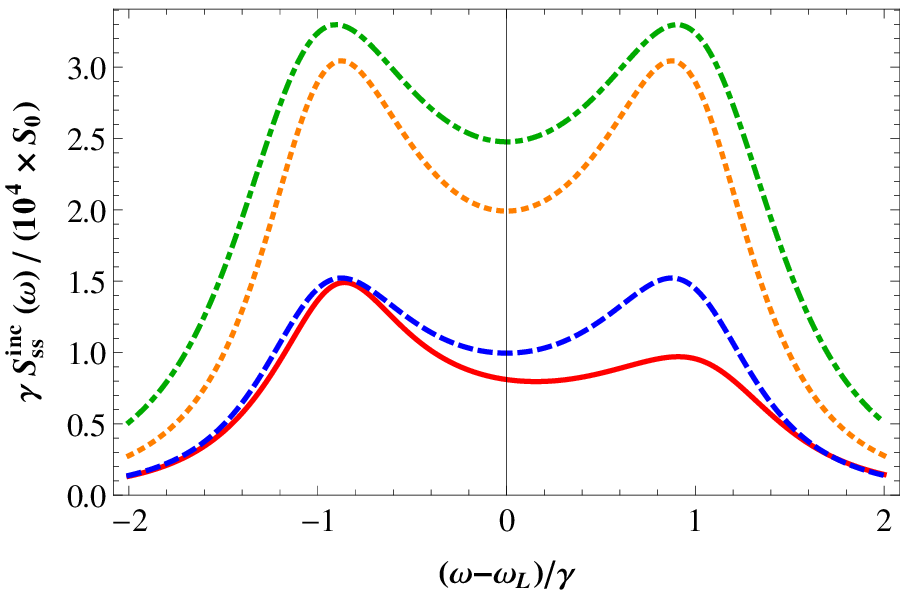}}\hspace{1cm}
\subfigure[{}]{\label{twoatomspecb}\includegraphics[width=8cm]{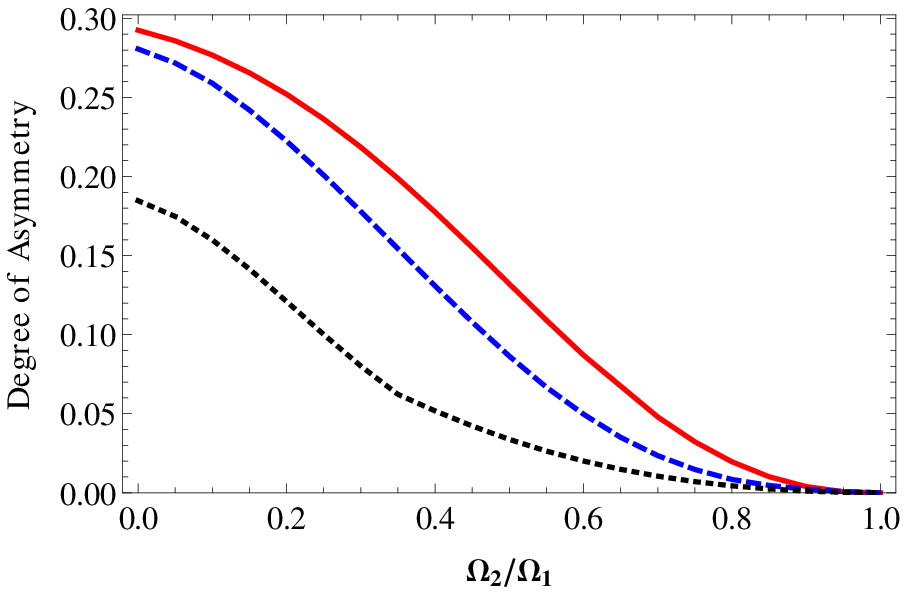}}
\caption{{\bf(a)} (Color online) Incoherent part of the power spectrum for 2 atoms 
with $\textbf{R}_{21}=a_l\vec{e}_x=0.82\lambda_0\vec{e}_x$, $\alpha^{12}=\pi/2$, $J^ {12}\approx-0.09\gamma$, 
$\Gamma^{12}\approx-0.11\gamma$, $\Delta=\gamma$, $\sigma=600$ nm and $\Omega_1=0.1\gamma$. The detector is 
positioned in the xy-plane with $\theta=\measuredangle(\hat{\textbf{r}},\textbf{R}_{21})\approx0.92$. The 
solid red and dashed blue line display the spectrum under single atom addressing, i.e. $\Omega_2\approx0.004\Omega_1\approx0$, 
while the dotted-dashed green and dotted orange line consider a broad laser beam with $0.1\gamma=\Omega_1\approx\Omega_2$. 
The dashed blue and the dotted orange line, however, display the case, where the collective parameters 
$J^{\mu\nu}$ and $\Gamma^{\mu\nu}$ have been 
artificially set to zero. \\
{\bf(b)} The degree of asymmetry as defined in Eq.~(\ref{eq:DOA}) 
plotted as a function of the ratio $\Omega_2/\Omega_1$ for parameters as in (a)
except for $\Omega_1=0.1\gamma\approx|J^{12}|\approx|\Gamma^{12}|$ (solid red line), 
$\Omega_1=0.5\gamma\approx 5|J^{12}|\approx5|\Gamma^{12}|$ 
(dashed blue line) and $\Omega_1=1\gamma\approx10|J^{12}|\approx10|\Gamma^{12}|$ (dotted black line).}
\label{2atomspec}
\end{figure*}
By making use of Eq.~(\ref{sourceterm}) the power spectrum can be expressed in terms of atomic 
two-time correlation functions,
\begin{equation}\label{eq:PowSS}
\begin{split}
S_{ss}(\textbf{r},\omega)=&S_0(\textbf{r}) \lim_{t\rightarrow\infty}\sum_{\mu,\nu=1}^N
\Re \left\{\int_{0}^{\infty} d\tau \;e^{-i(\omega-\omega_L)\tau}\right.\\
& \left.\;\;\left\langle\sigma_\mu^{+}(t+\tau)\sigma_\nu^{-}(t)\right\rangle\;
e^{ik_0\hat{\textbf{r}}\cdotp(\textbf{r}_\mu-\textbf{r}_\nu)}\right\},
\end{split}
\end{equation}
where $\Re\{\cdotp\}$ denotes the real part and 
$S_0(\textbf{r})=\frac{I_0 (\textbf{r})}{\pi}=\frac{1}{\pi}
\arrowvert \frac{\omega_0^2 \hat{\textbf{r}}\times(\hat{\textbf{r}}\times\vec{d}_0)}
{4\pi\varepsilon_{0}c^2r}\arrowvert^2$ 
denotes a normalization factor which contains the radiation properties of a dipole. 
The factor $e^{i\omega_L\tau}$ arises from the fact that we work in a rotating frame 
with respect to the operator $\frac{\hbar\omega_L}{2}\sum_{\mu=1}^N \sigma_\mu^z$. All the atomic 
one- and two-time correlation functions are accessible via the master Eq.~(\ref{eq:master}) 
and by usage of the quantum regression theorem \cite{QRT}. 

To analyze the power spectrum $S_{ss}$ as given in Eq.~(\ref{eq:PowSS}), it is helpful to split the 
expression into a coherent and an incoherent part. For two operators $\hat A$ and $\hat B$ the expectation 
value of the product $\hat A \hat B$ can always be separated into a coherent and an incoherent part \cite{Mollow,Glauber},
\begin{equation}
\Bra \hat A \hat B \Ket=\Bra \hat A\Ket \Bra\hat B \Ket+\left<(\hat A-\Bra \hat A\Ket)(\hat B-\Bra\hat B\Ket)\right>,
\end{equation}
respectively. It is therefore clear that the coherent part of the power spectrum is always proportional 
to a $\delta$-function, since
\begin{equation}
\begin{split}
S_{ss}^{co}(\textbf{r},&\omega)\propto\lim_{t\rightarrow\infty}\int_{-\infty}^{\infty}d\tau\; 
e^{-i(\omega-\omega_L)\tau}
\left\langle\sigma_\mu^{+}(t+\tau)\right\rangle \left\langle \sigma_\nu^{-}(t)\right\rangle\\
&= \Bra \sigma_\mu^+ \Ket_{ss}\Bra \sigma_\nu^- \Ket_{ss}\int_{-\infty}^{\infty}d\tau\; 
e^{-i(\omega-\omega_L)\tau}\propto \delta(\omega-\omega_L).
\end{split}
\end{equation}
Another feature of the power spectrum for $N\geq2$ atoms is the geometry dependence arising from the exponential
 $e^{ik_0\hat{\textbf{r}}\cdotp(\textbf{r}_\mu-\textbf{r}_\nu)}$. For atoms where the collective parameters 
$\Gamma^{\mu\nu}$ and $J^{\mu\nu}$ are negligible, this interference effect, however, does not contribute 
to the interesting part of the spectrum, namely the incoherent part. In this case the expectation 
values of $\Bra \sigma_\mu^+\sigma_\nu^-\Ket$ always factorize for $\mu\neq\nu$ and the interference 
terms only enter the coherent part of the spectrum. 
This results in an incoherent part of the spectrum which is the sum of single atom Mollow spectra. 
In cases where $\Gamma^{\mu\nu}$ and $J^{\mu\nu}$ are not negligible, the interference terms enter 
the incoherent part of the spectrum and can make it asymmetric. This effect can serve as a 
signature of single site addressing and lies at the center of our investigations.

\section{Breakdown of spectral symmetry} \label{results}

\subsection{Two Atoms}

To explain the basic physical mechanisms behind our results, it is convenient to concentrate on the spectrum 
of two atoms first. The position of atom 1 defines the point of origin, i.e. $\textbf{r}_1=\{0,0,0\}^\top$. 
Atom 2 is just positioned in the next lattice site at $\textbf{r}_2=\{a_l,0,0\}^\top$. For the lattice 
constant and the FWHM of the laser beam we choose $a_l=640$nm and $\sigma=600$nm like in Ref.~\cite{greiner}. The 
dipole moments are oriented along the z-axis, $\vec{d}_0\propto\vec{e}_z$, which leads to $\alpha^{12}=\pi/2$. For the 
D2 transition of Rubidium 87 (transition wavelength $\lambda_0=780$nm) the collective parameters result 
in $\Gamma^{12}\approx-0.11\gamma$ and $J^{12}\approx-0.09\gamma$.

Figure~\ref{2atomspec}(a) compares the case of a broad laser beam, i.e. $\Omega_1=\Omega_2=0.1\gamma$, to the 
case of a laser beam with a FWHM of $\sigma=600$nm focused onto atom 1, which results in 
$\Omega_1=0.1\gamma\approx25\Omega_2$. We refer to the latter case as single site addressing. It is seen that 
the spectral symmetry breaks down in the case of single site addressing under the effect of the 
dipole-dipole interaction.
For two atoms that are illuminated with equal intensity we find a spectrum of symmetric Mollow shape
as shown in Fig.~\ref{2atomspec}(a), see dotted-dashed green line.
Under single site addressing, however, the peak on the right hand 
side of the spectrum is suppressed which leads to an asymmetry, see solid red line.
To illustrate that this effect can be attributed to the presence of the dipole-dipole interaction $J^{12}$ 
and the collective damping rate $\Gamma^{12}$ we 
compare our results to the physically rather impossible situation where the collective parameters $\Gamma^{12}$ 
and $J^{12}$ are turned off artificially. As expected we find the symmetric Mollow spectrum; dashed blue curve for 
one atom being illuminated and dotted orange curve for two atoms. The latter only shows little deviations of 
the peak heights and positions from the case of $\Omega_{1} = \Omega_{2}$ and $J^{12}\neq0$, $\Gamma^{12}\neq0$. 

Note that only two peaks out of the 
triplet (at $\pm\Delta$) can be seen because of the weak driving strength $\Omega_0=0.1\gamma$, with $\Omega_0$ as 
defined in Eq.~(\ref{eq:gaussian_beam}). We choose such a weak Rabi frequency because the effect of symmetry 
breaking in the spectrum is more evident if the Rabi frequency is of the order of the dipole-dipole interaction. 
In the case where $\Omega_0\gg J^{\mu\nu}$ the dipole-dipole interaction is just a small perturbation to the driving 
of a single two-level atom, hence the symmetric Mollow shape dominates the spectrum.

Figure~\ref{2atomspec}(b) displays the degree of asymmetry in the spectra for different ratios 
of $\frac{\Omega_2}{\Omega_1}$ which is equivalent to different FWHM of the laser beam. 
We consider here a degree of asymmetry which is defined as
\begin{equation}\label{eq:DOA}
\begin{split}
D=
\frac{1}{S_{max}}\cdot
&\left(\max\{|S(\tilde{\omega})-S(-\tilde{\omega})|:\tilde{\omega}=\omega-\omega_L>0\}\right),
\end{split}
\end{equation}
normalized to the highest intensity $S_{max}$ in the spectrum,
\begin{equation}\label{eq:NormDOA}
S_{max}=\max\{S(\tilde{\omega})\;:\;\tilde{\omega}=\omega-\omega_L \in\Re\}.
\end{equation}
For parameters as in Fig.~\ref{2atomspec}(a) the definition of the degree of asymmetry corresponds
to the visibility of the difference in the peak heights. One can see that the degree of asymmetry decreases 
monotonically with increasing ratios of $\frac{\Omega_2}{\Omega_1}$ and goes exactly to zero as
$\Omega_2\rightarrow\Omega_1$. Furthermore the degree of asymmetry is highest if the Rabi frequency 
is comparable to the dipole-dipole interaction $J^{12}$ between the atoms.

In conclusion, the degree of asymmetry in the spectrum is a signature for the degree of single site 
addressing, provided that
the collective parameters $J^{\mu\nu}$ and $\Gamma^{\mu\nu}$ are of the order of the driving strength. 
Therefore we emphasize that single atom addressing in the regime of large inter-atomic separations is certainly 
not challenging but for small inter-atomic separations the collective atomic properties reveal whether single 
site addressing occurs in the system.

Our results also give direct evidence that the dipole-dipole interaction $J^{\mu\nu}$ is induced by 
mutual exchange of photons between the atoms. Indeed, as already mentioned the incoherent part of the 
spectrum only contains the interference of light emitted from separate atoms if the dipole-dipole 
interaction $J^{\mu\nu}$ is finite. Yet the interference effect is still present even if atom 2 is 
not illuminated by the laser at all, i.e. $\Omega_2=0$. This indicates that processes where atom 1 
absorbs a photon from the laser field, emits a photon which travels to atom 2 where it is absorbed and 
emitted again must exist. Since we look at the spectrum in the steady state limit and the Rabi frequency 
$\Omega_0$ is comparable to $J^{12}$ these processes occur at a rate large enough to generate the 
observed interference in the spectral light of the atomic ensemble. Note that this interference is not due 
to coherent classical light but it is rather of quantum mechanical nature.

\subsection{Larger numbers of atoms}

\begin{figure*}[htbp]$\begin{array}{lr}
\subfigure[{}]{\label{twoatomspeczta}\includegraphics[width=8cm]{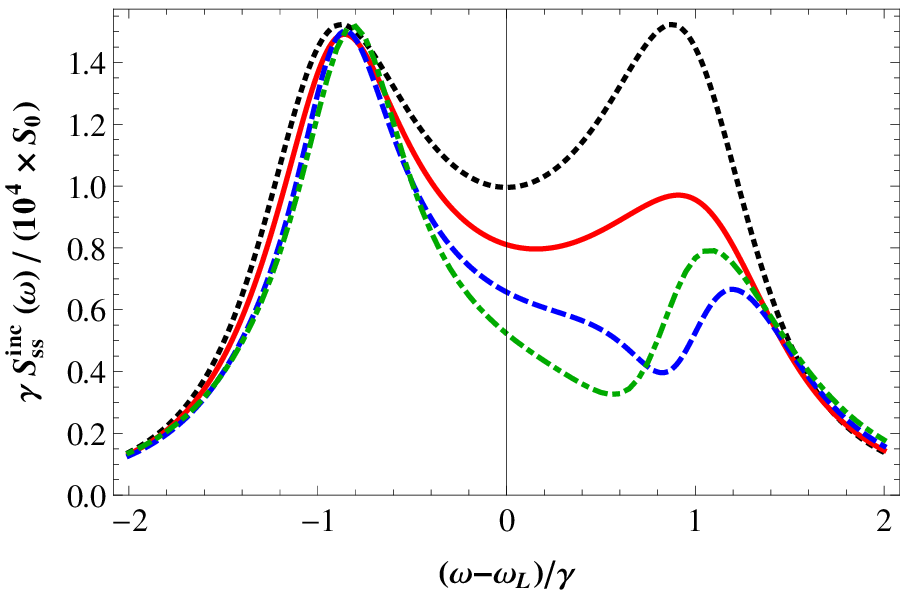}}\hspace{1cm}&
\subfigure[{}]{\label{twoatomspecztb}\includegraphics[width=8cm]{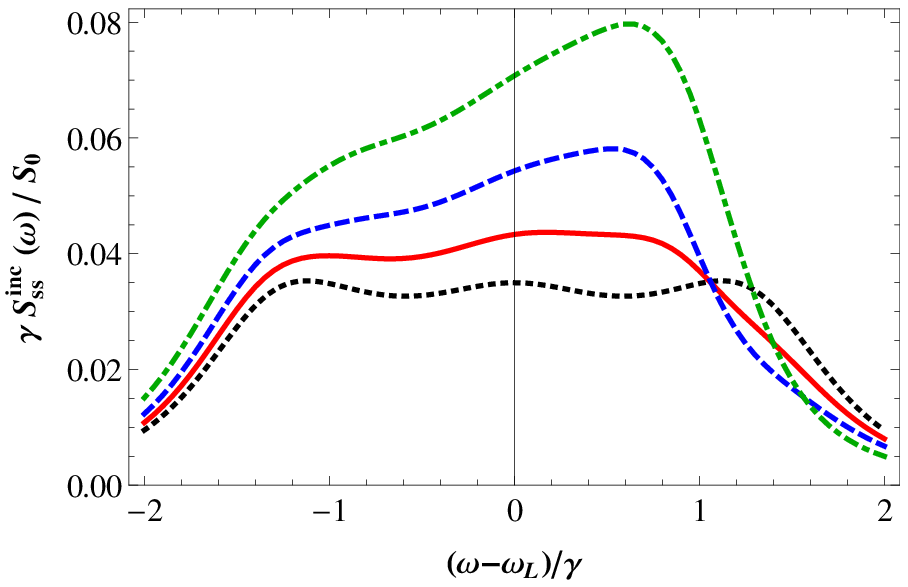}}\\
\subfigure[{}]{\label{twoatomspecztc}\includegraphics[width=8cm]{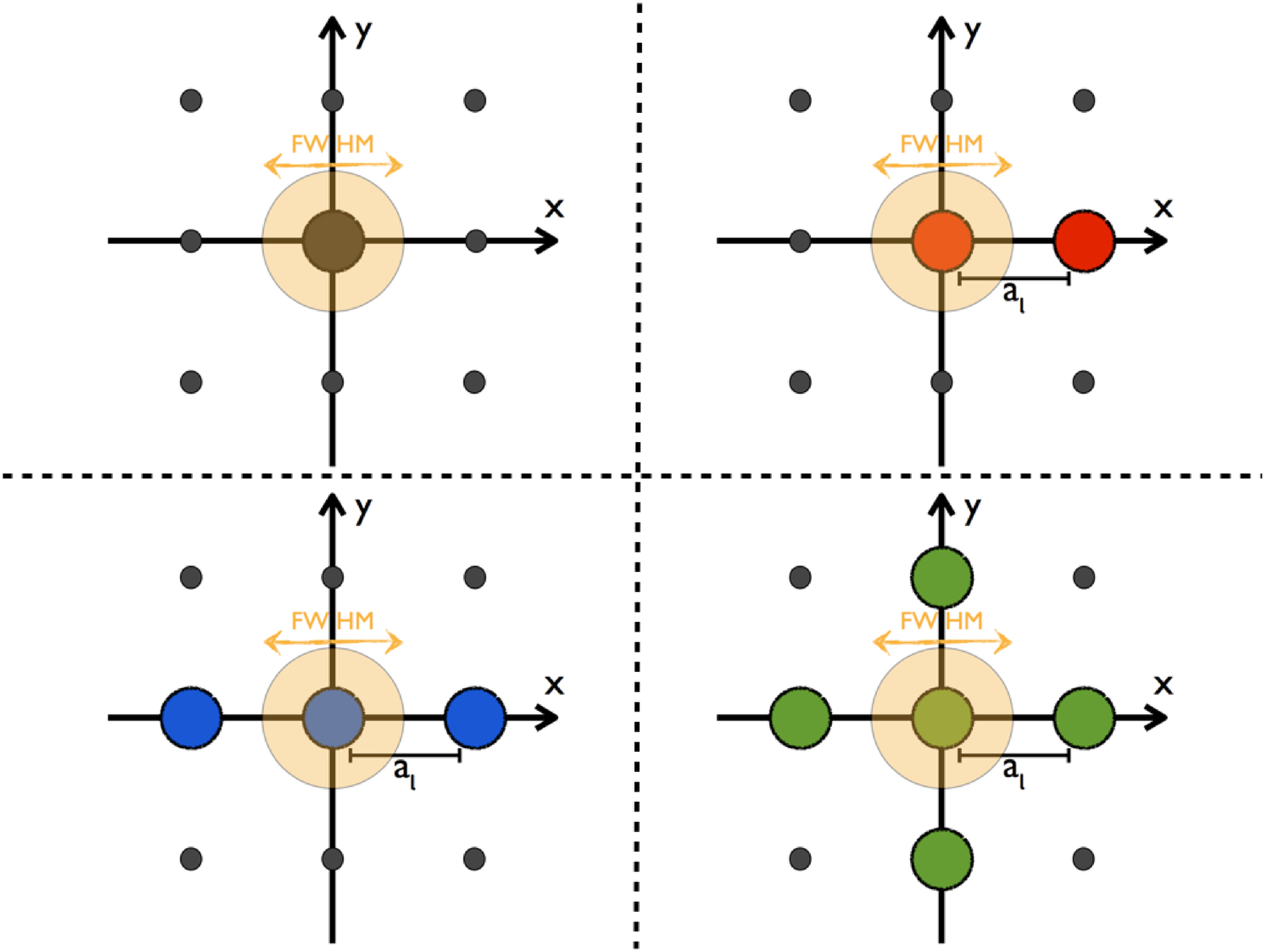}}&
\subfigure[{}]{\label{twoatomspecztd}\includegraphics[width=8cm]{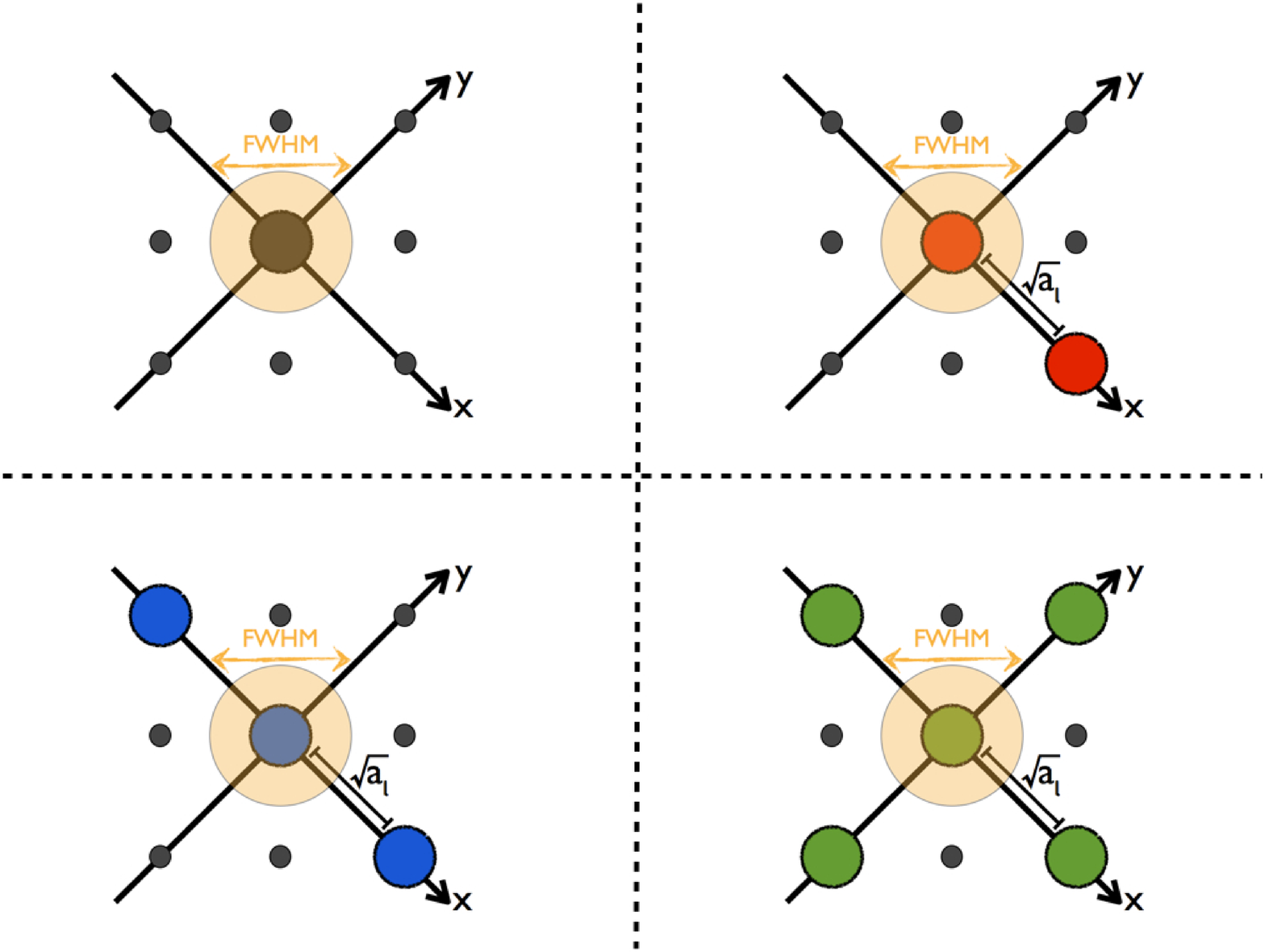}}
\end{array}$
\caption{(Color online) Incoherent part of the power spectrum (a), (b) with the corresponding 
atomic configurations in the lattice below, respectively (c), (d). 
{\bf(a)} Incoherent part of the power spectrum with parameters as in Fig.~\ref{2atomspec}(a). 
The solid red curve shows the spectrum for 2 atoms under single atom
addressing just as the solid red line in Fig.~\ref{2atomspec}(a). 
The dashed blue curve displays the spectrum of three atoms where the third atom is placed 
at $\textbf{r}_{3}=-a_l\vec{e}_x$.
This represents single site addressing in a 1D optical lattice where the contributions 
of the nearest neighbours of 
the addressed atom are taken into account. For the 2D lattice we place two more atoms 
at $\textbf{r}_{4}=a_l\vec{e}_y$ 
and $\textbf{r}_{5}=-a_l\vec{e}_y$ and find very similar features (dotted-dashed green curve).
The Mollow spectrum of a single atom is given also for comparison (dotted black curve).
{\bf(b)} Incoherent part of the power spectrum for 2 atoms (solid red curve), 
3 atoms (dashed blue curve), 5 atoms 
(dotted-dashed green curve) with the parameters $\Delta=\gamma$ and $\Omega_0=0.5\gamma$. 
The positions of the atoms are 
$\textbf{r}_{1}=\{0,0,0\}^\top$, $\textbf{r}_{2}=a_l\{\sqrt{2},0,0\}^\top$, 
$\textbf{r}_{3}=a_l\{-\sqrt{2},0,0\}^\top$, $\textbf{r}_{4}=a_l\{0,\sqrt{2},0\}^\top$ and 
$\textbf{r}_{5}=a_l\{0,-\sqrt{2},0\}^\top$, as illustrated in the picture below, 
with $a_l=532$ nm and $\sigma=700$ nm as 
in Ref. \cite{bloch,bloch2}. The detector is positioned in the far-field zone 
at $\hat{\textbf{r}}=\hat{r}\{0,1,0\}$. All the lines consider single site addressing which is 
compared to the Mollow line (dotted black curve). In {\bf(c)} and {\bf(d)}, small black dots indicate
lattice sites, colored dots indicate the atoms considered for plots of the corresponding color 
in (a) respectively (b) and the faint orange circle indicates the FWHM of the probe laser.}
\label{2atomspeczweiterteil}
\end{figure*}
If one considers the investigation of single site addressing in a 1D or 2D optical lattice more then 
one neighboring atom should be taken into account. 
Since the dipole-dipole interaction and the collective damping rates decay as 
$\frac{1}{x^{\mu\nu}}$ with increasing separation $x^{\mu\nu}$, one can however
obtain estimates for larger lattices by just considering a limited number of lattice sites or rather atoms
contributing to the spectrum.
We study the effects of single site addressing in a 1D and 2D optical lattice 
up to a level where the atoms with the highest contribution to the spectrum are taken into account.
These atoms are the ones where the dipole-dipole interaction potential $J^{1\mu}$ between the 
$\mu$-th atom and the addressed atom is largest in magnitude. 

Figure~\ref{2atomspeczweiterteil} displays the impact of single site 
addressing in a 1D optical lattice (dashed blue curve) and a 2D optical lattice 
(dotted-dashed green curve) configuration within the approximation described above and 
compares it to the case of two atoms (solid red curve) and to the Mollow spectrum of a single atom
(black dotted curve).
In part (a) of Fig.~\ref{2atomspeczweiterteil} we have chosen parameters for the lattice constant 
and the FWHM of the laser beam like in \cite{greiner}, i.e. $a_l=640$nm and $\sigma=600$nm. For this
lattice constant the magnitude of the dipole-dipole interaction is largest between the addressed atom
positioned at $\textbf{r}_1=\{0,0,0\}^\top$ and the nearest neighbours in the lattice, hence the atoms
positioned at $\textbf{r}=\{\pm a_l,0,0\}^\top$ and $\textbf{r}=\{0,\pm a_l,0\}^\top$. The parameters 
$\Delta=\gamma$ and $\Omega_1=0.1\gamma$ coincide with the parameters chosen in Fig.~\ref{2atomspec}. 
Again the detector is 
positioned in the xy-plane with $\theta=\measuredangle(\hat{\textbf{r}},\textbf{R}_{21})\approx0.92$. 
Therefore the solid red line in Fig.~\ref{2atomspeczweiterteil}(a) is the same as the solid red
line in Fig.~\ref{2atomspec}(a). It shows the incoherent part of the output spectrum for the addressed
atom at $\textbf{r}_1=\{0,0,0\}^\top$ in the presence of another atom at $\textbf{r}=\{a_l,0,0\}^\top$.
The dashed blue line displays the same situation but with a third atom placed at $\textbf{r}=\{-a_l,0,0\}^\top$.
The dotted-dashed green line shows the spectrum if a fourth and a fifth atom are placed 
at $\textbf{r}=\{0,\pm a_l,0\}^\top$.
In all cases we find a broken spectral symmetry under single site addressing. The peak on the
right hand side is suppressed for all atomic configurations as compared to the single atom
Mollow spectrum. We also note that
spacial symmetry of the atomic setup does not lead to spectral symmetry of the emitted light.

Figure~\ref{2atomspeczweiterteil}(b) displays the same situation as Fig.~\ref{2atomspeczweiterteil}(a) 
but with lattice parameters as in the experiments of Ref.~\cite{bloch,bloch2}. 
The lattice constant in the experiment is equal to $532$ nm. 
This leads to a 
dipole-dipole interaction of $J^{\mu\nu}\approx0.03\gamma$ between atoms positioned in two adjacent 
lattice sites,
while atoms separated by $\sqrt{2} a_l$ have a dipole-dipole interaction of $J^{\mu\nu}\approx-0.12\gamma$.
Hence, in our numerical calculations we concentrate on the contributions from these 
atoms and neglect the contributions from the nearest neighbours of the addressed atom.
Although the breaking of spectral 
symmetry is displayed in Fig.~\ref{2atomspeczweiterteil}(b) as well, we notice 
that the overall intensity 
of the case with a finite dipole-dipole interaction is a bit larger than for a single atom. 
Figure \ref{2atomspeczweiterteil}(a) exhibits the opposite behaviour. 
This effect is mainly due to the different positioning of the detector in 
Fig.~\ref{2atomspeczweiterteil}(a) and \ref{2atomspeczweiterteil}(b).

If the atoms in the configurations of Fig.~\ref{2atomspeczweiterteil}(c) and Fig.~\ref{2atomspeczweiterteil}(d) are all
driven by the same driving strength we find symmetric power spectra in all cases. In the case of two atoms 
the degree of asymmetry is exactly zero. In the cases of three or five atoms we find a degree of
asymmetry which is negligible but not exactly zero. For two equally driven atoms
the master equation is fully symmetric under the exchange of these two atoms. This is not longer true in the case of
three or five equally driven atoms as for example $J^{12}\neq J^{23}$ because of the geometry 
dependence of the dipole-dipole interaction.
However, if we choose $J^{\mu\nu}\equiv J$ and $\Gamma^{\mu\nu}\equiv \Gamma$ for all $\mu,\nu \in\{1,...,N\}$ 
artificially, and drive all atoms with the same Rabi frequency,
we find a spectrum for 2,3,4 or 5 atoms with a degree of asymmetry that is exactly zero. 
In the next section we discuss this observation further. 

\section{Some conclusions and remarks about the symmetry of the
  spectra}\label{symmetry chapter}

Our results lead us to a remarkable first conclusion, regarding our
particular system:

\begin{itemize}
\item[1.] In the case where the master equation and therefore also the
  density matrix, is fully symmetric under the exchange of each
  possible pair of two-level atoms $\mu$ and $\nu$ (i.e. under atomic
  permutation), the total spectrum of emission of the system is
  symmetric around the laser frequency.
\end{itemize}

Moreover, putting together these results with many other examples one
could think of~\cite{elenathesis}, we further envision a second
conclusion regarding any open quantum system in general:

\begin{itemize}
\item[2.] Provided that a single quantum system, QS, (such as a
  few-level system or harmonic oscillator under coherent or incoherent
  continuous excitation) exhibits a symmetric steady state power
  spectrum, then a number $N$ of such QSs coupled to each other, in a
  way that the density matrix is fully symmetric under all possible
  permutations of these QSs, will also exhibit a symmetric total
  steady state power spectrum. Consequently all possible auto- and
  cross-correlation functions between pairs of QSs will be real and
  therefore experimentally observable.
\end{itemize}

The second conclusion is a generalization of the first one. These
statements are difficult to proof starting from the properties of the
Liouvillian of an ensemble of coupled QSs. Here we only explore some directions
that such a proof may take.

The power spectrum for $N$ QSs with associated operator
$\sigma_\mu$, consists of a sum of $N^2$ contributions
$S_{\mu\nu}(\tilde\omega)$, each given by
\begin{equation}\label{eq:Spec}
S_{\mu\nu}(\tilde\omega)\propto \Re \left\{\int_{0}^{\infty} d\tau\;
e^{-i\tilde\omega\tau}\langle\sigma_\mu^{+}(\tau)\sigma_\nu^{-}\rangle_{ss}\right\}\,,
\end{equation}
with $\langle\sigma_\mu^{+}(\tau)\sigma_\nu^{-}\rangle_{ss}
=\lim_{t\rightarrow\infty}\langle\sigma_\mu^{+}(t+\tau)\sigma_\nu^{-}(t)\rangle$
and $\tilde{\omega}=\omega-\omega_L$. Each of these terms can be
decomposed into a symmetric and an asymmetric part simply by
separating the corresponding correlator into its real and imaginary
parts:
\begin{equation}\label{eq:Specaufteilung}
\begin{split}
  S_{\mu\nu}(\tilde\omega)&=S^{\mathrm{sy}}_{\mu\nu}(\tilde\omega)+S^{\mathrm{asy}}_{\mu\nu}(\tilde\omega)=\\
  &\int_{0}^{\infty} d\tau\;
  \Re\{\langle\sigma_\mu^{+}(\tau)\sigma_\nu^{-}\rangle_{ss}\}\cos(\tilde\omega\tau)\\
  &-\int_{0}^{\infty} d\tau\;
  \Im\{\langle\sigma_\mu^{+}(\tau)\sigma_\nu^{-}\rangle_{ss}\}\sin(\tilde\omega\tau)\,.
\end{split}
\end{equation}
In the case where the master equation, and therefore also the density
matrix, is fully symmetric under QS permutation, we have
$S_{\mu\nu}(\tilde\omega)=S_{\nu\mu}(\tilde\omega)$. Consequently, the
total spectrum can be computed in terms of two different correlators
only, for example
\begin{equation}
  S(\tilde\omega)=N S_{11}(\tilde\omega)+N(N-1) S_{12}(\tilde\omega)\,.
\end{equation}
In this case the spectrum is symmetric if and only if every correlator
$\langle\sigma_\mu^{+}(\tau)\sigma_\nu^{-}\rangle_{ss}$ is real. Then,
the two-time operator $\sigma_\mu^{+}(\tau)\sigma_\nu^{-}$ with
$\mu\neq\nu$ becomes an observable in the steady state as we stated in
our second conclusion.

However, the fact that the correlators become real when the spectrum
is symmetric does not provide new information about the system, it is
only a mathematical implication. The general question when an open
quantum system should have a symmetric spectrum around some relevant
frequency (the laser in the case of coherent excitation) is not a
trivial one. First, this depends crucially on the nature of the excitation
that is being detected, that is, the operators appearing in the
two-time correlator. In our case generalized to~$N$ atoms, we refer to
the collective Dicke operator
$\sigma^{\pm}_\mathrm{S}=\sum_{\mu=1}^N\sigma_\mu^{\pm}$ but it could
be any deexcitation operator in the system that corresponds to some
physical entity. Each peak that appears in the spectrum is related to
the \emph{probability amplitude} to transit between two eigenstates of the
system by emitting one of these quasiparticles. This means that both
the dynamics of the dressed states and their quasiparticle component
play a role. In order to make this link clearer, let us decompose the
incoherent part of a spectrum into a sum of $d^2$
peaks~\cite{elenathesis},
\begin{equation}
  \label{eq:decompEle}
  S(\tilde\omega)=\frac{1}{\pi}\sum_{p=1}^{d^2}\Big[\frac{L_p\gamma_p/2-K_p(\tilde\omega-\omega_p)}{(\gamma_p/2)^2+(\tilde\omega-\omega_p)^2}\Big]\,,
\end{equation}
with $\omega_p$, $\gamma_p$ (peak position and linewidth), $L_p$ and
$K_p$ (Lorentzian and dispersive weights) all real parameters and~$d$
the dimension of the Hilbert space. Then, $-(i\omega_p+\gamma_p/2)$
are the eigenvalues of the Liouvillian in matrix form, $\mathbf{L}$, which are
either real (giving rise to a single Lorentzian peak at the center) or
pairs of complex conjugates (giving rise to a pair of sister peaks
symmetrically placed around the center, with equal broadening)~\cite{footnote1}.

A given pair of
sister peaks (with $i\omega_\alpha+\gamma_\alpha/2=-i\omega_\beta+\gamma_\beta/2$) is
symmetric if the complex weights are also conjugates,
\begin{equation}\label{liovillianbalance}
L_\alpha+iK_\alpha=L_\beta-iK_\beta
\end{equation}
These are computed from the eigenvectors
of~$\mathbf{L}$ including the steady state density matrix. They correspond
exactly to the transition probability amplitude as mentioned above.
The balance between two amplitudes give rise to a
pair of symmetric twin peaks. If this is the case for all sister
peaks, the total spectrum is, of course, symmetric.

In order to grasp
all the physical sense of this balance condition, one would need 
to identify (or rather reconstruct) the eigenstates of the system
under study. 
This is not an easy task, especially in presence of
both dissipation and excitation~\cite{elenafermions,elenaqubits}. In
some limiting cases, however, such as one strongly driven two-level
system, it is possible~\cite{cohen,detailedbalance}. The eigenvectors of the full
Liouvillian correspond in good approximation to the so-called
\emph{dressed states}, $\ket{\pm}$, obtained from diagonalizing the
Hamiltonian part only. As in this regime the three peaks that form the
Mollow structure are well separated, the interference part of the
spectrum is negligible ($K_p\approx 0$) and the condition for symmetry
of the two side peaks, $L_\alpha=L_\beta$, is completely equivalent to the
so-called \emph{detailed balance} between the dressed states:
$\rho_{++}\mathcal{P}_{(+\rightarrow
  -)}=\rho_{--}\mathcal{P}_{(-\rightarrow +)}$, where $\rho_{\pm\pm}$
are the dressed state steady state populations and
$\mathcal{P}_{(\pm\rightarrow \mp)}$ the transition rates between
them. On the other hand, if the excitation is weak and there is an
overlap between the peaks of the spectrum the Hamiltonian, dressed
states are not anymore the eigenstates of the full Liouvillian and
their detailed balance is no longer a necessary condition for symmetry
(in fact it breaks down out of resonance where the spectrum is still
symmetric).

A symmetric spectrum thus implies that the probability amplitudes 
of transitions between eigenstates are balanced. However, the challenging
task of reconstructing such eigenstates makes it difficult to foresee
and demonstrate when a system will exhibit a symmetric
spectrum. Specially in a configuration where coupling strength, decay
and excitation rates are of the same order of magnitude, as in this
study. 
Putting Eq.~(\ref{liovillianbalance}) in terms of the eigenvectors of the Liouvillian
gives the mathematical condition that they must fulfill so that the spectrum is symmetric.
But this does not bring any further
insight into the matter if one cannot identify which properties of the system
cause the weights $K_\alpha$ and $L_\alpha$ of the corresponding
Liouvillian $\mathbf{L}$ to fulfill this equation.

On the other hand, our conclusions 1 and 2
are somehow intuitive and expected if one reasons on physical grounds,
so let us end this Section with a plausible explanation for the
symmetry of the spectrum and its break down in our particular case.

In our configuration, the QS is simply a two-level atom driven by a
laser, whose Mollow spectrum is indeed always symmetric (in the
absence of incoherent pumping, pure dephasing or other decoherence
effects). When assembling $N$ of such identical and identically driven
QSs, new collective states are expected to appear. The driving in this case
is restricted to the set of states 
$\ket{0}\leftrightarrow \sigma_S^+\ket{0}\leftrightarrow...\leftrightarrow (\sigma_S^+)^N\ket{0}$
that form an ($N+1$)-level system.
The remaining non-symmetric states are not driven at all but provide an effective decay channel for
the ($N+1$)-level system. It is known from dressed state arguments that the total spectrum of a
coherently driven ($N+1$)-level system is symmetric. 
Additional effective decay through the remaining non-symmetric states
does not break the symmetry of the total spectrum, as we calculated in previous sections. 
Now, if the atoms are not driven equally the laser does not solely drive the ($N+1$)-level system
but also the transitions between non-symmetric states. 
This disrupts the dynamics of the ($N+1$)-level system and induces
decoherence in the form of pure dephasing and an effective incoherent pump. Both
elements are well known to break the symmetry in the spectra of
coherently driven systems~\cite{Agarwal,Elena2}. Letting other atomic
parameters be different, such as decay rates or detunings, has similar
decoherent impact on the dynamics and the symmetry of the spectrum.
We have checked that this is the case for systems consisting of up to
five two-level atoms.

\section{experimental applicability} \label{exp}

Our numerical results in section \ref{results} clearly show the signatures of single site addressing in resonance fluorescence spectra.
In summary it can be stated that the spectrum is symmetric around the laser frequency if the atoms are illuminated
by a laser with equal strength. If, in contrast, only one atom is addressed the spectrum becomes asymmetric.
Thus it is possible to measure a fluorescence spectrum and deduce information about the quality of an addressing scheme.

In optical lattices a large number of ultra cold atoms are trapped. As soon as these atoms are illuminated by light the atoms heat up unavoidably. 
The detection methods presented in the experiments \cite{bloch,bloch2,greiner} are all destructive measurements in the sense that the atomic sample
is too hot after detecting the atoms. The single site addressing scheme presented in Ref.~\cite{bloch2} is designed in a way that the test for the failure or success
of the addressing scheme requires a destructive measurement. Hence the test and the usage of the addressing scheme have to be carried out in different
atomic samples.
 
Yet, if one makes a resonance fluorescence measurement as pictured in this work only a small fraction of atoms will be heated. Atoms
positioned at the edges of a 2D lattice, for example, are well suited to test the addressing laser. After the test of the addressing scheme
the laser can be moved over the lattice to the desired position. The test and the usage of the addressing scheme could be carried out in only
one atomic sample.

\section{conclusion} \label{conclusionsattheend}

In this work we calculated and investigated the resonance fluorescence spectra of two-level atoms under the 
influence of local addressability. We implemented local addressing by means of a laser beam focused on a single 
atom in an atomic ensemble. A master equation in a Markovian regime modeled the interaction between the atoms and 
the surrounding electromagnetic vacuum. Due to small inter-atomic separations we had to account for a dipole-dipole 
interaction between atoms that is induced by the mutual exchange of photons and collective damping processes.
With numerical calculations we demonstrated that the output power spectrum of an atomic ensemble is 
asymmetric in the case of single atom addressing. We showed that this effect is generated by the 
presence of the dipole-dipole interaction due to photon exchange.
Our results suggest that resonance fluorescence measurements could provide sensitive tests for the 
addressing of individual atoms in 1D or 2D optical lattices.
They thus also allow to predict emission spectra of 1D optical lattices on the surface of optical
nanofibres~\cite{fibre}.
Our calculations are valid for any set of
two-level quantum systems. The applicability of 
resonance fluorescence measurements as a test for single site addressability is therefore not restricted
to neutral atoms in optical lattices but also applies to quantum dots~\cite{dots} and colour centers in diamonds~\cite{NVcenters}.
Furthermore, we provided some physical and intuitive explanation for the symmetry of the spectrum and
its breakdown under single site addressing. We finally generalized our findings as follows:   
Provided that a single quantum system exhibits a symmetric steady state power
spectrum, this property also holds for $N$ identical and identically coupled quantum systems.

\acknowledgements

The authors thank Martin Kiffner for fruitful discussions at an earlier stage of this project.
This work is part of the Emmy Noether project HA 5593/1-1 funded by the DFG and was
supported by the Alexander von Humboldt foundation and the DFG-CRC 631.


\begin{thebibliography}{99}
\bibitem{footnote1}
Because the densitiy matrix is hermitian, in vectorial 
representation, $\vec{\rho}^*=T\vec{\rho}$ 
where~$T$ is some similarity transformation. 
Therefore $\mathbf{L}$ and $\mathbf{L}^*=T\mathbf{L}T$ have the same characteristic polynomial
with real coefficients.
\end{thebibliography}
\end{document}